\documentclass{article}
\usepackage{spconf,amsmath,graphicx,verbatim,subfig}
\usepackage{mathtools}
\usepackage{makecell}
\usepackage{amsfonts}
\usepackage{setspace}
\usepackage{url}
\usepackage{booktabs}
\usepackage{cite}
\usepackage{hyperref}
\usepackage{multirow}

\title{HappyQuokka System for ICASSP 2023 Auditory EEG Challenge}
\name{Zhenyu Piao\quad Miseul Kim\quad Hyungchan Yoon\quad Hong-Goo Kang}
\address{Department of Electrical and Electronic Engineering, Yonsei University, Seoul, South Korea}

\begin{document}
% \ninept
%
\maketitle

\begin{abstract}
\vspace{-6pt}
% 1.Work for {Official challenge name}
This report describes our submission to Task 2 of the Auditory EEG Decoding Challenge at ICASSP 2023 Signal Processing Grand Challenge (SPGC).
Task 2 is a regression problem that focuses on reconstructing a speech envelope from an EEG signal.
% 2.Proposed Framework
For the task, we propose a pre-layer normalized feedforward transformer (FFT) architecture.
For within-subjects generation, we additionally utilize an auxiliary global conditioner which provides our model with additional information about seen individuals.
% Specifically,
% 3.Results
Experimental results show that our proposed method outperforms the VLAAI baseline and all other submitted systems.
Notably, it demonstrates significant improvements on the within-subjects task, likely thanks to our use of the auxiliary global conditioner.
In terms of evaluation metrics set by the challenge, we obtain Pearson correlation values of 0.1895 $\pm$ 0.0869 for the within-subjects generation test and 0.0976 $\pm$ 0.0444 for the heldout-subjects test.
% 특별히, within-in subject task에서 뛰어난 성능을 보였다.
We release the training code for our model online.\footnote{Source code for our system is available in \normalfont{\url{https://github.com/jkyunnng/HappyQuokka_system_for_EEG_Challenge}}.}
\end{abstract}

%keyword 추가완료
\vspace{-12pt}
\section{Introduction}
\label{Introduction}

Task 2 of the Auditory EEG Challenge hosted by the 2023 ICASSP Signal Processing Grand Challenge (SPGC) Committee is a regression task that focuses on establishing a relationship between measured electroencephalography (EEG) signals and an auditory stimulus (speech envelope).
The goal of the task is to reconstruct the auditory stimuli given the EEG signals.
To do this, we designed a Transformer-based feed-forward network \cite{fastspeech} with pre-layer normalization \cite{prelayernorm}. 
We additionally utilize an auxiliary global conditioner \cite{oord2016wavenet} to provide the specific subject information to the system.
Our model, which is trained by negative Pearson correlation with L1 loss, demonstrates significantly better results than the VLAAI model \cite{vlaai} and other competing systems.
Specifically, it obtains a correlation of 0.1895 $\pm$ 0.0869 for the within-subjects task and 0.0976 $\pm$ 0.0444 for the heldout-subjects task in terms of the challenge's evaluation metrics. 
In the following sections, we describe the design of our system in greater detail.

% Figures
\begin{figure}
\centering
\subfloat[Model overview]{\includegraphics[width=0.25\textwidth]{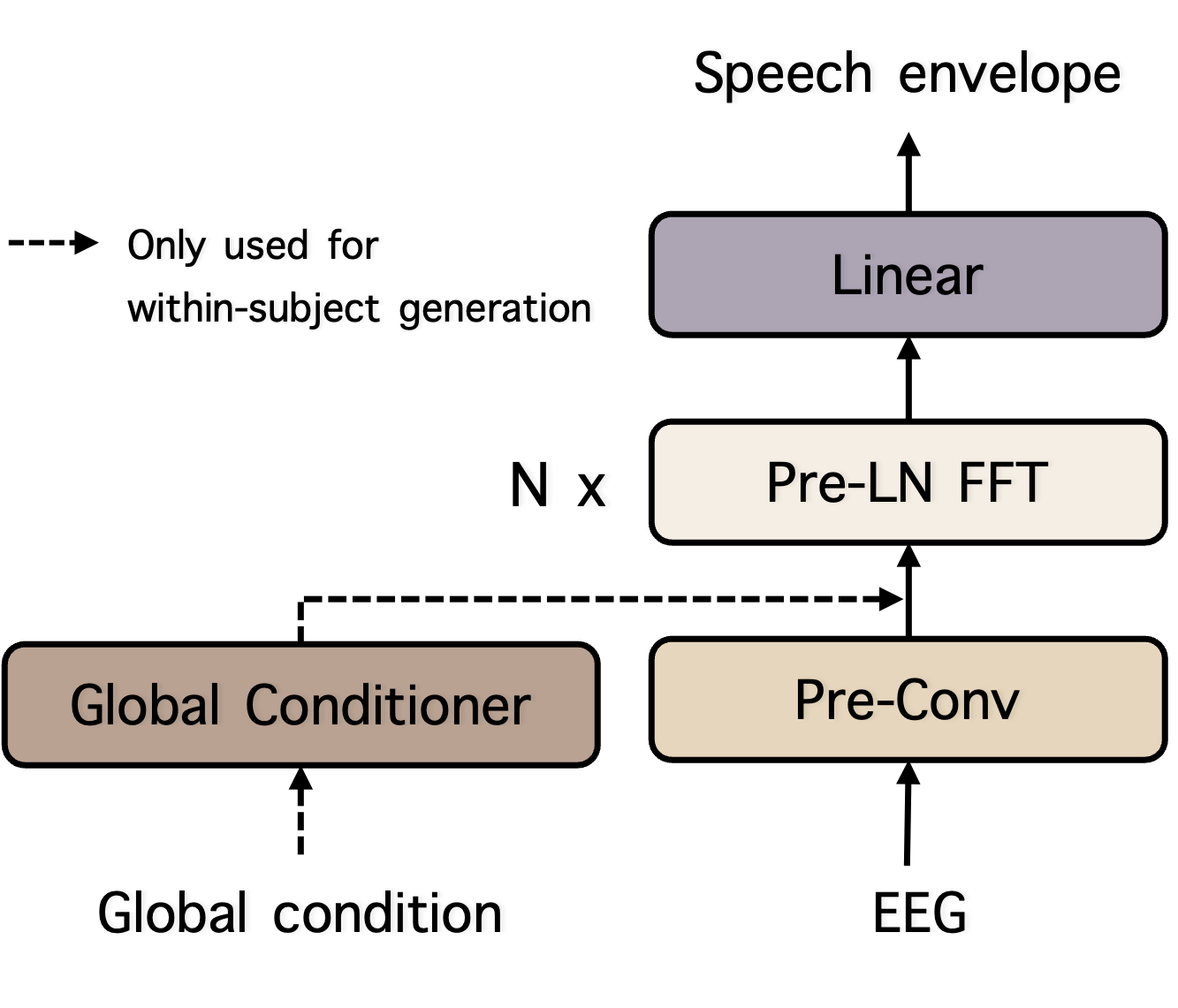}}
\hspace*{15pt}  
\subfloat[Pre-LN FFT]{\includegraphics[width=0.12\textwidth]{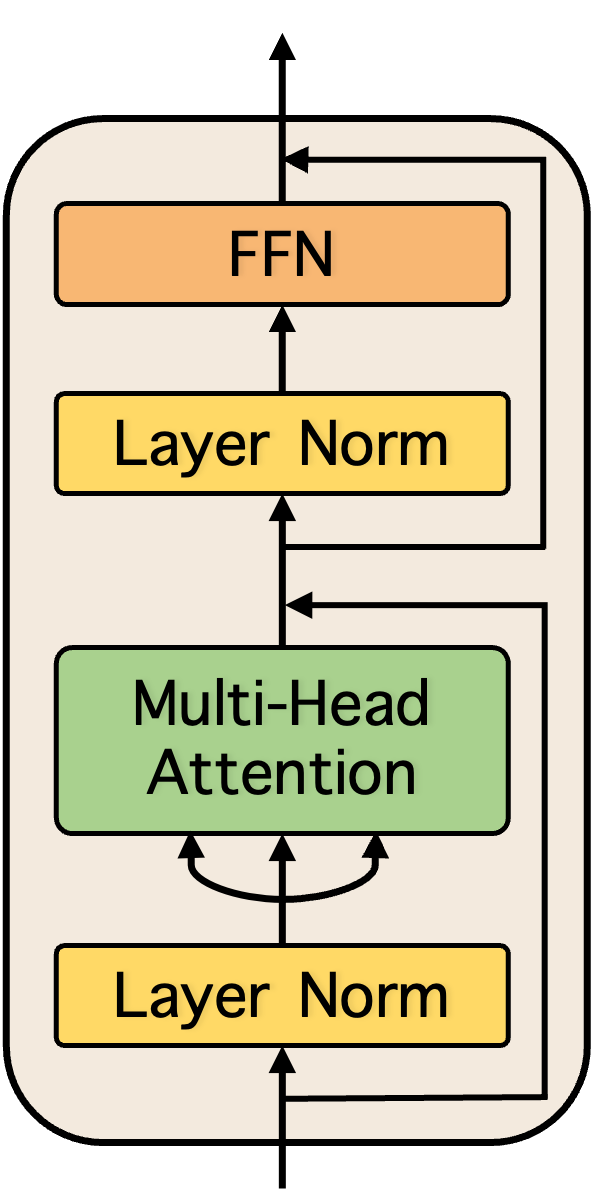}}
\caption{(a) Overview of the proposed model and (b) detailed architecture of the Pre-LN FFT block.
Note that the global conditioner is only used for the within-subjects generation.} 
\label{fig:1}
\vspace{-6pt}
\end{figure}
\vspace{-6pt}

\section{Model}
\label{Model}
\vspace{-8pt}
% proposed model
% 1.transformer-based architecture with pre-layer normalization
% 2.auxiliary global conditioner (provide subject information to the system as condition)
\vspace{-4pt}
\subsection{Overview}
The backbone architecture of our proposed model is a Feed-Forward Transformer (FFT) network\cite{fastspeech} with some modifications.
The model consists of a pre-conv layer, several FFT blocks with pre-layer normalization (pre-LN) \cite{prelayernorm}, an auxiliary global conditioner, and a linear layer (Figure \ref{fig:1}).
The input EEG features are fed into the pre-conv layer, then are passed through the pre-LN FFT blocks to model the relationship between the EEG signal and the speech stimulus using the Transformer's self-attention mechanism. 
The output latent vector is passed through a linear layer to summarize this information into a single speech envelope. 
For within-subjects generation, we additionally design an auxiliary global conditioner to provide the subject identity as an extra input.

\vspace{-15pt}
\subsection{Pre-LN FFT block}
The architecture of a standard FFT block consists of a feed-forward structure, which utilizes the Transformer's self-attention mechanism and 1D convolutions. 
For our model, we make some slight modifications to standard FFT blocks by placing the layer normalization inside the residual blocks instead of putting it between the residual blocks \cite{prelayernorm}.
We call this module a pre-LN FFT block.

\vspace{-4pt}
\subsection{Auxiliary global conditioner}
The auxiliary global conditioner is applied for within-subjects generation, and provides additional information about each subject to the model.
The auxiliary global conditioner takes a one-hot encoding of a subject's identity and expands it into a subject embedding that has the same dimension as the output of the pre-conv layer.
The two embeddings are added together and fed into the pre-LN FFT blocks.
By conditioning our model on additional global input variables,
we are able to guide it to generate the speech envelopes of seen individuals more accurately by taking into account their different brain activity characteristics. 

\vspace{-4pt}
\subsection{Loss function}
For the loss function, in addition to the original negative Pearson loss $R$, we also add $L_{1}$ loss, which minimizes the absolute differences between predicted and target values to ensure stable training.
Thus, the final training loss is given as:
\begin{equation}
L_{total} = -R + \mathcal{\alpha} *
L_1.
\end{equation}
Here, we experimentally set the optimal value of $\alpha$ to 0.2.
\vspace{-6pt}
\section{Experiments}
\label{Experiment}

\vspace{-6pt}
\subsection{Dataset}
We used the official dataset \cite{eegdata_K3VSND_2023} provided by the challenge organizers to train and test our model.
Here, EEG singals are sampled in 64 Hz.
We directly utilized the pre-processed data that was already normalized and split into train, validation, and test subsets in a 1:1:1 ratio.
For training, we used the train subset, which contains 508 different EEG-speech pairs. For model evaluation, we used the test subset, which also contains 508 data of within-subjects generation.

\vspace{-6pt}
\subsection{Implementation details}
We implemented our proposed model and VLAAI baseline using PyTorch. 
The number of pre-LN FFT layers was set to 8, and we used 2 heads for multi-head attention.
We trained our model, as well as all baseline models, for 1000 epochs using the Adam optimizer with an initial learning rate of 0.0005.
We also applied a StepLR scheduler with a learning rate decay factor of 0.9.
During training, we used 5-second segments of signals for stable training, which was randomly cropped from each EEG/speech envelope segment.
For inference, we split input signals into several 5-second-long segments, then concatenated the outputs to make the whole envelope. 
%to calculate the correlation, then calculate the mean of those correlation values as final metric.

\vspace{-6pt}
\subsection{Results}
% Table 1 summarizes all the experimental results of the baseline and proposed network.
%0.vlaai
%1. prenorm
% 2. global condition
% 3.loss
Table \ref{table1} summarizes the results of the VLAAI baseline and our proposed model, as well as several variants that we trained for ablation studies.
It is shown that our model outperforms the VLAAI baseline by a large margin.
Our ablation studies also demonstrate the impact of the various modules of our network on its overall performance.
Finally, in terms of the challenge's evaluation metrics, 
for within-subjects generation, we obtained a correlation value of 0.1895 $\pm$ 0.0869,
while for the heldout-subjects task, we obtained 0.0976 $\pm$ 0.0444.

% Table 1
\begin{table}[]
\begin{spacing}{0.62}
\caption{Pearson correlation values of the VLAAI baseline and our proposed model, as well as ablations.
% Note that the results are delineated in regards to within-subject generation.
}
% And the relative performance improvement of our systems compared with the VLAAI baseline.}
\resizebox{\linewidth}{!}{%
\begin{tabular}{ccc}
\toprule
Model & \begin{tabular}[c]{@{}c@{}}Pearson correlation\end{tabular} & \begin{tabular}[c]{@{}c@{}}Performance\\ improvement (\%)\end{tabular} \\ \midrule
Baseline (VLAAI)       & 0.1614 & - \\
Proposed               & \textbf{0.2029} & \textbf{25.7} \\ \midrule
w/o pre-LN           & 0.1896 & 17.5 \\
w/o global conditioner & 0.1829 & 13.3 \\
w/o L1 loss            & 0.1965 & 21.8 \\
\bottomrule
\end{tabular}%
}
\label{table1}
\end{spacing}
\vspace{-6pt}
\end{table}
\vspace{-6pt}

% \vspace{-6pt}
\section{Conclusion}
\label{Conclusion}
This report describes our system for the Auditory EEG Decoding Challenge at ICASSP 2023. Our proposed model is composed of pre-LN FFT blocks and an auxiliary global conditioner, which is especially beneficial for within-subjects generation. 
Experimental results show that our proposed model outperforms the baseline VLAAI system provided by the organizers as well as other submitted systems. 
\section{Acknowledgement}
\label{Acknowledgement}
\vspace{-8pt}

This work was supported by the project `Alchemist Brain to X (B2X) Project' through the Ministry of Trade, Industry and Energy (MOTIE), South Korea, under Grant 20012355 and NTIS 1415181023. 
\vspace{-6pt}

% \clearpage
\bibliographystyle{IEEEtran}
\bibliography{ref}

\end{document}